\documentclass[lettersize,journal]{IEEEtran}
\usepackage{amsmath,amsfonts}
\usepackage{amssymb}
\usepackage{algorithmic}
\usepackage{algorithm}
\usepackage{array}
\usepackage[caption=false,font=normalsize,labelfont=sf,textfont=sf]{subfig}
\usepackage{textcomp}
\usepackage{stfloats}
\usepackage{url}
\usepackage{verbatim}
\usepackage{graphicx}
\usepackage{cite}
\usepackage{bm}     
\newtheorem{theorem}{Theorem}

\newtheorem{proposition}{Proposition}

\usepackage[svgnames,dvipsnames]{xcolor}   
\usepackage[colorlinks=true, allbordercolors=white]{hyperref} 

\hypersetup{
    linkcolor=blue,    
    citecolor=blue,  
}

\begin{document}

\title{Performance Bounds for Near-Field Velocity Estimation With Modular Linear Array}

\author{Khalid A. Alshumayri,
        Mudassir Masood, \IEEEmembership{Senior Member, IEEE},
        Ali A. Nasir, \IEEEmembership{Senior Member, IEEE}%
\thanks{Khalid A. Alshumayri is with the Department of Electrical Engineering,
King Fahd University of Petroleum and Minerals, Dhahran 31261, Saudi Arabia.}%
\thanks{Mudassir Masood and Ali A. Nasir are with the Electrical Engineering
Department and the Interdisciplinary Research Center for Communication Systems
and Sensing (IRC-CSS), King Fahd University of Petroleum and Minerals,
Dhahran 31261, Saudi Arabia.}}



\maketitle

\begin{abstract}
Velocity estimation is a cornerstone of the recently introduced near-field predictive beamforming. This paper derives the Cramér–Rao bounds (CRBs) for joint radial and transverse velocity estimation within a predictive beamforming framework employing a modular linear array (MLA). We obtain approximated closed-form CRB expressions that characterize the interplay between array geometry and estimation accuracy, showing that increasing the inter-module separation enlarges the effective aperture and reduces the transverse-velocity CRB, while the radial-velocity CRB remains largely insensitive to this separation. Furthermore, we show that an MLA can achieve the same accuracy as a collocated array with fewer antennas and quantify the relation between inter-module spacing and antenna savings. The derived expressions are validated through simulations by comparing them with the mean-squared error (MSE) of the maximum likelihood estimator (MLE) reported in the literature. 
\end{abstract}


\section{Introduction}
\IEEEPARstart{T}{he} adoption of large-aperture arrays has given rise to a near-field (NF) channel model that assumes a spherical rather than planar wavefront. This makes the NF channel jointly dependent on the propagation direction and distance. Moreover, the curvature of the spherical wavefront encodes both radial and transverse motion, enabling the estimation of both velocity components using a monostatic array. Based on these estimated velocities, the target’s future direction and distance can be forecast. This is the key idea behind the predictive beamforming framework in \cite{predc2024}, where a maximum likelihood estimator (MLE) is employed for NF velocity sensing without requiring pilot signals. However, this pioneering work did not investigate fundamental performance bounds for velocity estimation. Reference \cite{crb2024giov} derived the Cramér–Rao bounds (CRBs) for both radial and transverse velocities for single-input multiple-output (SIMO) monostatic radar, which differs from the predictive-beamforming framework considered here.

\begin{figure}[t]
  \centering
  \includegraphics[width=0.8\linewidth]{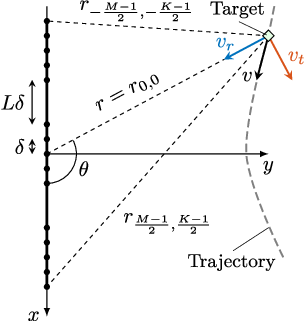}
  \caption{Near-field velocity sensing employing a modular linear array.}
  \label{fig:mod_array}
\end{figure}

Realizing beamforming with a large number of antennas poses significant challenges in both hardware and signal processing. To mitigate this issue, the modular linear array (MLA) architecture, which consists of multiple collocated uniform linear array (ULA) modules, enables the construction of a larger effective aperture. As illustrated in Fig.~\ref{fig:mod_array}, the modules are separated by inter-module spacing larger than the inter-element spacing, easing deployment on rooftops or building façades while extending the NF region \cite{Emil_modular,mod_multiple_AoA,mod_beam_focusing_Li,mod_first_Li}. Prior work on MLA has derived the CRBs for range and angle estimation. Specifically, \cite{crb_mod} derives Cramér–Rao bounds (CRBs) for near-field bistatic sensing in widely-spaced multi-subarray MIMO, and quantifies how the array geometry and the TX–RX distance affect the best-achievable parameter-estimation accuracy under both a spherical-wave model and a hybrid spherical/planar-wave model. This analysis was extended in \cite{crb_mod_nonuniform} to MLAs with non-uniform inter-module spacing. However, CRBs for near-field velocity estimation with MLAs have not been reported in the literature. 

In this letter, we derive the performance bound for joint velocity estimation using MLA. The contributions of this work are: (i) We derive approximated CRB expressions for joint radial and transverse velocity estimation in the NF region, revealing explicit dependence on the parameters of MLA, and verify that the mean-squared error (MSE) of the MLE in \cite{predc2024} is efficient and attains the derived CRBs. (ii) We obtain a closed-form expression for the predictive-beamforming array gain under velocity mismatch due to estimation errors and show that transverse-velocity estimation error contributes more severely to performance degradation compared to radial-velocity counterparts. (iii) We establish a practical design rule: enlarging inter-module separation reduces the transverse-velocity CRB via effective-aperture growth, while the radial-velocity CRB remains essentially unchanged. This suggests that the MLA architecture can achieve the same transverse-velocity estimation accuracy as a collocated ULA, but with fewer antennas. However, reducing the total number of antennas incurs only a negligible increase in the radial-velocity CRB.

\section{Signal and System Model}

As illustrated in Fig.~\ref{fig:mod_array}, we consider a near-field communications system. The base station (BS) is equipped with a modular antenna array composed of $K$ modules, each with $M$ antennas, resulting in a total of $MK$ antennas. The antenna spacing is uniform and denoted by $\delta$. The modules are placed evenly along the $x$-axis with an inter-module spacing $L\delta$. The location of the $m$-th element in the $k$-th module is given by the vector $\mathbf{w}_{m,k} = [x_{m,k}, 0]^T$ where $m=0,\pm1, \cdots, \pm \frac{M-1}{2}, \ k=0,\pm1, \cdots , \pm \frac{K-1}{2}$. The coordinates are given by $x_{m,k} = (Uk+m)\delta$, where $U = M+L-1$. Further, the same antenna array is employed for transmitting and receiving, made possible by circulators and full-duplex capability \cite{full_duplex}. The $(m,k)$-th entry of the array response vector at time index $n$ is given by \cite{zhaolin_mod_to_pro, predc2024}

\begin{equation}
    \big[ \mathbf{a}_n(r,\theta,v_r,v_t) \big]_{m,k} = \exp\left\{-j\frac{2\pi}{\lambda}(r_{m,k} + v_{m,k}\ nT_s)\right\} \, , \label{eq:array_response}
\end{equation}

\noindent where $r_{m,k} = \sqrt{r^2 - 2r(Uk+m)\delta \cos\theta + (Uk+m)^2\delta^2}$ is the distance between the target and the $(m,k$)-th antenna in module $k$, and $T_s$ is the symbol duration. The Doppler shift is caused by $v_{m,k} = q_{m,k} v_r + p_{m,k} v_t$, where $v_r$ and $v_t$ denote the radial and transverse velocities of the target with respect to the origin, $q_{m,k}$ and $p_{m,k}$ denote the projection coefficients

\begin{align}
q_{m,k} &= \frac{r - (Uk+m)\delta \cos\theta}{r_{m,k}}, p_{m,k} = \frac{(Uk+m) \delta \sin\theta}{r_{m,k}} \label{projection}
\end{align}

\noindent The coefficients satisfy $q_{m,k}^2 + p_{m,k}^2 = 1$, and if the FF model is used, such that $r_{m,k} \approx r - (Uk+m)\delta\cos\theta$, it will result in $q_{m,k} = 1$ and $p_{m,k} = 0$, which explains the reason the transverse velocity remains unobservable under that model. Besides, we assume a point-like moving target whose motion remains in the radiating NF region, at a distance greater than the Fresnel distance \cite{tut_near}, which is defined as $d_F = 0.5\sqrt{ A^3/\lambda }$ where $A = \delta[K(M-1)+L(K-1)]$ denotes the aperture length, which can equivalently be expressed as $A=\delta \left[U(K-1)+M-1\right]$. 

For notation convenience, let $\bm{\phi} $ represent the target parameters $\{r,\theta, v_r, v_t \}$. Assuming that the downlink channel between BS and target is dominated by the line-of-sight (LoS) component of the channel, the received baseband echo signal at the array, reflected by the moving target, at time index $n$ for the $l$-th coherent processing interval (CPI) will be described as 

\begin{equation}
    \mathbf{y}_l(n) = \beta  \ \mathbf{a}_n(\bm{\phi}_l)\mathbf{a}_n^\text{T}(\bm{\phi}_l) \mathbf{x}(n) + \mathbf{z}(n)
    \label{signalModel}
\end{equation}

\noindent where $\beta \in \mathbb{C}$ represents the reflection coefficient that depends on both the round-trip pathloss and the radar cross section (RCS), $\mathbf{z}(n)$ is circular white Gaussian noise. The transmit signal $\mathbf{x}(n)$ is designed according to the estimated parameters $\bm{\phi}$ from the previous CPI. 
\subsection{Predictive Beamforming}
The advantages of employing predictive beamforming in designing the transmit signal $\mathbf{x}(t)$ are twofold: 1) Doppler compensation and, and 2) prediction of the target's future location by utilizing the kinematic model
\begin{align}
    \hat{r}_{l+1} = \hat{r}_l + \hat{v}_{r, l} \ NT_s \, , \ \
    \hat{\theta}_{l+1} = \hat{\theta_l} + \hat{v}_{t,l} \ NT_s / \hat{r}_l \, .
    \label{kin_model}
\end{align}
\noindent The predicted parameters $\bm{\phi}$ are used to steer the beam in the next CPI. Accordingly, the transmit signal is designed as  
\begin{equation}
    \mathbf{x}(n) = \frac{\mathbf{a}_n^*(\hat{\bm{\phi}_{l}})}{\sqrt{MK}} s(n)
    \label{PreCoding}
\end{equation}

\noindent where $s(n)$ represents the information symbol that satisfy $|s(n)|^2=1$. A rough estimate of the parameters $v_r$ and $v_t$ can be obtained from the previous CPI and used in the current CPI to compensate for the Doppler shift and predict the target's future location \cite{predc2024}. Substituting \eqref{PreCoding} into \eqref{signalModel}, the received echo signal at the BS will be \footnote{A similar derivation applies to the uplink scenario. If the transmitted signal is known (e.g., a pilot), then $\mathbf{y}_l(n)=\beta \ \mathbf{a}_n(\bm{\phi}_l) s(n)$.}

\begin{equation}
    \mathbf{y}_l(n) = \beta \ \psi(n) \ \mathbf{a}_n(\bm{\phi}_{l}) \ s(n) + \mathbf{z}(n) \, ,
    \label{eq:SignalModel}
\end{equation}

\noindent where $\psi(n)$ is the array-gain function. It characterizes the array gain under velocity mismatch due to estimation error from the previous CPI 

\begin{align}
\psi(n) & = \frac{1}{\sqrt{MK}}\sum_{m,k} e^{-j\frac{2\pi}{\lambda} \left( q_{m,k}\Delta_{v_r} + p_{m,k}\Delta_{v_t}\right)nT_s } \, ,
\label{psi} 
\end{align}

\noindent where $\Delta_{v_r}$ and $\Delta_{v_t}$ denote the mismatches between the true and estimated radial and transverse velocities, respectively. 

\subsection{Array-Gain Function}

To simplify $\psi(n)$, $p_{m,k}$ is approximated as $p_{m,k}\approx (Uk+m)\delta\sin\theta/r$ for aperture lengths satisfying $A\gg\lambda$, as established in Theorem \ref{Theorem}. Furthermore, since $q_{m,k} \gg p_{m,k}$, we can approximate it as $q_{m,k}\approx1$. Hence, $\psi(n)$, $p_{m,k}$ becomes

\begin{align}
    \psi(n) & \approx \frac{e^{-j\frac{2\pi}{\lambda}\Delta_{v_r} nT_s}}{\sqrt{MK}}  \left(  \sum_{m} e^{-j\tilde{\Delta}_{v_t} m}\right) \left(\sum_{k} e^{-j\tilde{\Delta}_{v_t} Uk} \right) \, ,
\end{align}
\noindent where $\tilde{\Delta}_{v_t} = \tfrac{2\pi}{\lambda} \left(\frac{\delta\sin\theta}{r}\right) \Delta_{v_t} nT_s$. The summation can be recognized as a Dirichlet kernel function. Hence, it directly follows that

\begin{align}
    \psi(n) = \frac{e^{-j\frac{2\pi}{\lambda}\Delta_{v_r} nT_s}}{\sqrt{MK}} \frac{\sin(M\tilde{\Delta}_{v_t}/2)}{\sin(\tilde{\Delta}_{v_t}/2)} \frac{\sin(KU\tilde{\Delta}_{v_t}/2)}{\sin(U\tilde{\Delta}{v_t}/2)} \, .  \label{psi_simplified}
\end{align}

\noindent Ideally, when $\hat{v}_r \approx v_r$ and $\hat{v}_t \approx v_t$, the array gain satisfies $\psi(n) \approx \sqrt{MK}$, which is numerically demonstrated in Section~\ref{sec:results}. Besides, applying the point scatterer model, the reflection coefficient can be expressed as \cite{braun2014ofdm}, $|\beta |^2 = \frac{P_t G_tG_r \lambda^2\sigma_{RCS}}{ (4\pi)^3 \ r^4} \,$, where $P_t$ and $\sigma_{RCS}$ represent the transmit power and the radar cross-section, respectively, while, $G_t$ and $G_r$ denote the antenna gains for transmitter and receiver respectively. For a monostatic setup, the antenna gains are equal, i.e., $G_r = G_t$. 

\section{CRBs for Velocity Estimation}

The CRB provides a theoretical lower limit on the variance of unbiased estimators of a deterministic parameter. It is derived from the Fisher Information, which quantifies the amount of information that an observable random variable carries about an unknown parameter. The Fisher information matrix (FIM) for radial and transverse velocity is given by

\[
\mathbf{F} = \begin{bmatrix}
J_{v_rv_r} & J_{v_rv_t} \\
J_{v_tv_r} & J_{v_tv_t}
\end{bmatrix}\,.
\]
\noindent Let the unknown parameters $\bm{\zeta} = [v_r \ v_t]^T$, the FIM elements under the white Gaussian noise model can be found as  \cite[Ch.~15.7]{GOAT}

\begin{align}
    J_{ij} = \frac{2}{\sigma^2} \sum_{n} \sum_{m} \sum_{k} \Re\bigg\{ \frac{\partial \mathbf{u}^{H}(n)}{\partial \zeta_i}\frac{\partial \mathbf{u}(n)}{\partial \zeta_j} \bigg\}\, , \nonumber
\end{align}

\noindent where $\frac{\partial}{\partial \zeta}\mathbf{u}(n)= \beta \sqrt{MK} \ \frac{\partial}{\partial \zeta}\mathbf{a}_n(\bm{\phi}) \ s(n) $ is the noise-free version of the received echo signal at the BS. Accordingly, the three FIM elements can be readily obtained $J_{v_rv_r} = \gamma MK \sum_{m,k}^{} q^2_{m,k}$, $J_{v_tv_t}  = \gamma MK \sum_{m,k}^{} p^2_{m,k}$, and $J_{v_rv_t}  = \gamma MK \sum_{m,k}^{}q_{m,k} \ p_{m,k}$. where we define the SNR as

\begin{align}
    \gamma = \bigg(\frac{2\pi}{\lambda}\bigg)^2 \frac{|\beta|^2N(N+1)(2N+1)}{3\sigma^2} T^2_s \, . 
\end{align}

\noindent Then, the CRBs for radial and transverse velocities correspond to the diagonal elements of the inverse of the FIM.

\begin{align}   
    \text{CRB}(v_r) = \frac{J_{v_tv_t}}{\text{Det}\ \mathbf{F}} \, ,\
    \text{CRB}(v_t) = \frac{J_{v_rv_r}}{\text{Det}\ \mathbf{F}} \, .
    \label{exact_crb}
\end{align}

\noindent To obtain more tractable CRB expressions, we employ a simple yet tight approximation. In particular, FIM can be simplified using the following theorem

\begin{theorem}\label{Theorem}
Consider a target located at a distance \( r \ge d_F \), and assume that the aperture satisfies
$\delta\left[ U(K-1) + M - 1\right] \gg \lambda$. Then the following approximations hold:
\begin{align}
    \sum_{m,k} p_{m,k}^2 
        &\approx \frac{MK}{12} \big( U^2 (K^2-1) + (M^2-1) \big)
           \frac{\delta^2}{r^2}\sin^2\theta , \label{sum:pm,k} \\
    \sum_{m,k} q_{m,k} p_{m,k} 
        &\approx \frac{MK}{12} \big( U^2 (K^2-1) + (M^2-1) \big)
           \frac{\delta^2}{r^2}\cos\theta \sin\theta . 
\end{align}
\end{theorem}

\begin{IEEEproof}
Please refer to Appendix \ref{Appendix_A}.
\end{IEEEproof}

\noindent For a standard uniform linear array, $\lambda/\delta = 2$, hence the condition $U(K-1)+M-1\gg2$ is typically satisfied. The expression \eqref{sum:pm,k} shows that the information on transverse velocity increases with the inter-module separation through the factor $U^2$. Moreover, we can derive the closed-form CRBs as follows

\begin{align}
    \text{CRB}^{(v_r)}_{\text{Mod}} &= \frac{12}{\gamma (MK)^2 \left(12 - \left(\frac{\delta}{r}\right)^2(U^2(K^2-1)+M^2-1) \right)}\, , \label{eq:derived_crb_vr} \\
    \text{CRB}^{(v_t)}_{\text{Mod}} &= \frac{12(r/\delta)^2}{\gamma (MK)^2(U^2(K^2-1)+M^2-1) \sin^2 \theta} \, .  \label{eq:derived_crb_vt}
\end{align}

\noindent The existence of the CRBs is guaranteed when $\frac{12}{U^2(K^2-1) +M^2-1} > (\delta/r)^2$, which holds for $U(K-1)+M-1\gg\lambda/\delta$. Note that the radial CRB is approximately not dependent on the target angle because the projection component $q_{m,k}$ is insensitive to the target angle since the quantity $\delta/r_{m,k}$ is relatively small. Also, it is largely unaffected by the inter-module spacing owing to the small term $(\delta/r)^2$. However, increasing the inter-module spacing reduces the transverse CRB. Additionally, for the special case when $L=1$ and $M_{0}=MK$, the result is reduced to the CRB for the conventional ULA in \cite{crb2024giov}

\begin{align}
    \text{CRB}^{(v_r)}_{\text{ULA}} &= 
    \frac{12}{\gamma M_{0}^2 \left(12 - \left(\frac{\delta}{r}\right)^2(M_{0}^2-1) \right)}\, , \label{eq: derived_crb_vr_single} \\
    \text{CRB}^{(v_t)}_{\text{ULA}} &= 
    \frac{12 (r^2/\delta^2)}{\gamma M_{0}^2(M_{0}^2-1) \sin^2 \theta} \, .  \label{eq: derived_crb_vt_single}
\end{align} 

\begin{proposition} \label{proposition} Since $\delta/r \ll 1$, the radial CRB is much less than the transverse CRB, as confirmed numerically in Section~\ref{sec:results}. Because the increased inter-module spacing results in $\text{CRB}^{(v_t)}_{\text{Mod}} < \text{CRB}^{(v_t)}_{\text{ULA}}$, it is natural to ask whether we can save some antennas by widening the spacing while reducing the number of elements, so that the MLA transverse CRB matches the collocated ULA transverse CRB. We have shown through analysis below and later through numerical results that with an MLA having inter-module separation equivalent to $25\%$ of the aperture of a   collocated ULA, we can remove approximately $17\%$ of the antennas in the ULA while achieving the same transverse-velocity CRB.
\end{proposition}

\begin{IEEEproof}
Formally, we seek parameters for which
\begin{align}
    \text{CRB}^{(v_t)}_{\text{ULA}} = \text{CRB}^{(v_t)}_{\text{Mod}} .
    \label{eq:crb_match}
\end{align}

\noindent Concretely, for some choice of inter-module separation $\bar{L}$ and number of antenna per-module $\bar{M}$ satisfying $K\bar{M}<M_o$, the modular array’s CRB will equal that of the ULA. To this end, we parameterize the inter-module spacing (in multiples of $\delta$) as $\bar{L} = 1 + \eta(M_{0}-1)$, where $\eta$ denotes the additional spacing between adjacent modules expressed as a fraction of the reference $M_0$-element ULA aperture. The minimum separation that achieves the same performance as ULA can be directly obtained from \eqref{eq:crb_match} and is given by

\begin{align}
    \eta =  \frac{1}{M_{0}-1} \left( \sqrt{\frac{M_{0}^2(M_{0}^2-1)-(\bar{M}^2-1)(\bar{M}K)^2}{(K^2-1)(\bar{M}K)^2}}  - \bar{M} \right) \, . \label{eq:separation_eta}
\end{align}

\noindent The spacing $\eta$ can be written in terms of the fraction of antennas used $h =K\bar{M}/M_0$, assuming $M_0\gg1$

\begin{align}
    \eta = \sqrt{ \frac{K^2-h^4}{K^2(K^2-1)h^2}} - \frac{h}{K} \, .
    \label{design_rule}
\end{align}

\noindent Note that $\eta$ is an explicit function of $h$ and $K$. For example, if $h=0.825$ (i.e., $17.5\%$ antenna saving) and $K=2$, then $\eta=0.25$. This verifies Proposition \ref{proposition}, and we will verify it through numerical results in the next section.
\end{IEEEproof}

\section{Numerical Results} \label{sec:results}

This section presents numerical results validating the CRBs for radial and transverse velocity in the modular linear array. The carrier frequency is $28$~GHz, the bandwidth is $B=100$~kHz, and the CPI is $2$~ms, yielding a pulse width of $T_s=10^{-5}$~s and $N=200$ symbols per CPI. In addition, we set $G_t=G_r=1$ and $\sigma_{RCS}=-23$ dB. The noise density power is set to $-174$ dBm/Hz and the transmit power $P_t = -10$ dBm.

\begin{figure}[htbp]
  \centering
  \includegraphics[width=\linewidth]{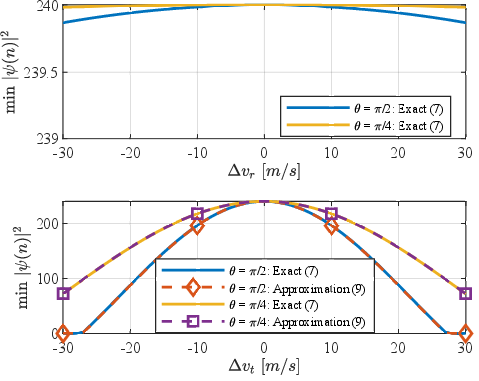}
  \caption{Effect of velocity mismatch on the array gain when $r=d_F$ and $L=61$.}
  \label{fig:Doppler_mismatch}
\end{figure}

Fig.~\ref{fig:Doppler_mismatch} shows the array-gain reduction due to velocity estimation errors. More specifically, the minimum (i.e., worst) value of $|\psi(n)|^2$ within a CPI is plotted against radial $\Delta v_r$ and transverse $\Delta v_t$ velocity mismatch. It is evident that transverse-velocity mismatch results in a much more pronounced array-gain reduction than radial-velocity mismatch. This increased sensitivity to $\Delta v_t$ arises because the transverse velocity induces an element-dependent Doppler shift, which damages coherent combining across the array elements. Furthermore, the closed-form array gain expression in \eqref{psi_simplified} closely matches the exact expression in \eqref{psi}. 

\begin{figure}[htbp]
  \centering
  \includegraphics[width=\linewidth]{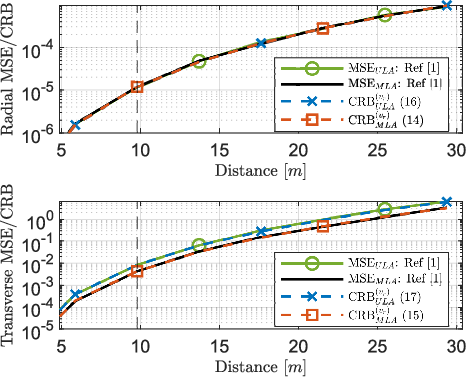}
  \caption{The MSE of MLE in \cite{predc2024} and the derived CRBs. The total number of antennas used is $240$. For the modular array, the number of modules is $2$, the number of antennas in each module is $M=120$, and the modular spacing parameter $L=61$.}
  \label{fig:mse}
\end{figure}

The mean-squared error (MSE) of the MLE proposed in \cite{predc2024} is plotted in Fig.~\ref{fig:mse} and compared to the derived CRBs. The MSE is estimated via Monte Carlo simulation with 1000 iterations for each value. The ground-truth velocities are $v_r=10$ m/s and $v_t=8$ m/s. In addition, the MLE is initialized with the values $(11, 7)$. Although this initialization is close to the ground-truth values, it is justified within the predictive beamforming framework \cite{predc2024}. It can be noted that the MSE closely matches the derived CRBs. Overall, the estimation accuracy of $v_r$ and $v_t$ differs substantially. The transverse CRB is much larger than the radial CRB, because the transverse velocity contributes significantly less Fisher information. In addition, the transverse CRB of the MLA is lower than that of the ULA because of MLA's larger aperture length, whereas the radial CRB is identical in the two cases. 

\begin{figure}[htbp]
  \centering
  \includegraphics[width=0.95\linewidth]{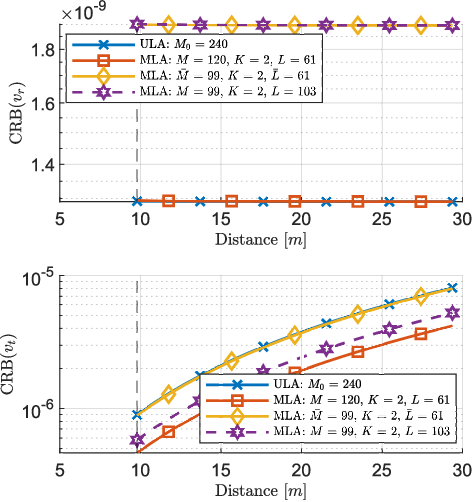}
    \caption{CRB for radial velocity (top) and transverse velocity (bottom), with $\theta=\pi/2$. The gray dashed line indicates the Fresnel distance of the MLA with $M=120$ and $L=61$.} 
  \label{fig:crb_reduced_number_of_antennas}
\end{figure}

To understand the interplay between array geometry and estimation accuracy,  Fig.~\ref{fig:crb_reduced_number_of_antennas} presents the radial and transverse CRBs for different array configurations without considering the effect of pathloss. The 240-element MLA achieves lower transverse CRB than collocated ULA, while the radial CRB is not affected by the inter-module separation. In addition, the first 198-element MLA design (solid, $\bar{L}=61$, $K=2$, and $\bar{M}=99$) matches the 240-element collocated ULA’s transverse-velocity CRB while using about $17.5\%$ fewer elements. However, the reduction in the number of antennas costs about $1.6$ dB increase in the radial CRB, which is relatively negligible. The second 198-element MLA design (dashed, $L=103$, $K=2$, and $\bar{M}=99$) has wider inter-module spacing to attain the same aperture as the 240-element MLA but with less number of antennas yet, the radial CRB is not affected. These comparisons confirm that enlarging inter-module separation reduces the transverse-velocity CRB through effective-aperture growth, whereas the radial-velocity CRB is largely insensitive to separation because the dependence enters only via the small squared ratio $(\delta/r)^2$ in~\eqref{eq:derived_crb_vr}. 

\section{Conclusion}
In this paper, we derive approximated CRBs for joint radial and transverse velocity estimation with MLAs in the NF regime and establish a closed-form characterization of the array gain in the presence of velocity mismatch. The analysis reveals that inaccurate transverse-velocity estimation can severely degrade the beamforming gain. Our results also show that increasing the inter-module separation enlarges the effective aperture and reduces the transverse-velocity CRB, whereas the radial-velocity CRB is largely insensitive to this separation. Leveraging this behavior, we derive an explicit relation between the inter-module spacing and the number of antennas required to meet the same transverse-velocity CRB as a collocated ULA. For example, with a two-module MLA and an inter-module separation equal to $25\%$ of a ULA aperture, the MLA matches the ULA's transverse-velocity accuracy while using $17\%$ fewer antennas. Future work could include adopting an extended-target model instead of a point-like target, as well as investigating velocity estimation in multi-target scenarios.


{
\appendix[Proof of Theorem 1] \label{Appendix_A}
To simplify the summation, $g=Uk+m$ where $\epsilon = \delta/r$, 

\begin{align}
   p_{m,k}^2 & = \frac{\big(g\epsilon \sin\theta\big)^2}{1 - 2(g\epsilon)\cos\theta+(g\epsilon)^2} \, .
    \label{pm_appendix}
\end{align}

\noindent Set $x = g \epsilon$ and define the function $f(x) \triangleq \frac{x^2\sin^2\theta}{1-2x\cos\theta+x^2}$. Observe that, $x \in \left[-\bar{x},\bar{x}\right]$, where the maximum value of $x$ occurs at the edge of the array i.e., $\bar{x}=\left(U\frac{K-1}{2}+\frac{M-1}{2}\right)\epsilon$. Given that $\epsilon\ll1$ and its maximum value occurs at the Fresnel distance when $\epsilon=\delta/d_F$, then


\begin{equation}
    \text{max}\{|\bar{x}|\} =  \sqrt{\frac{\lambda}{A}} \,,
\end{equation}


\noindent the quantity $|\bar{x}|$ can be made arbitrary small if and only if the aperture length $A = \delta [U(K-1)+M-1] \gg\lambda$. Then, using Taylor series expansion around $x=0$, one can approximate $f(x)\approx x^2 \sin^2\theta$. Since $p^2_{m,k} = f(g \epsilon) \approx (g \epsilon )^2 \sin^2\theta$, then the summation becomes

\begin{align}
    \sum_{m,k}{p^2_{m,k}} &\approx (\epsilon\sin\theta)^2  \sum_{m,k} (Uk+m)^2 \, ,\nonumber \\ &= (\epsilon\sin\theta)^2 \sum_k \sum_{m}\left( U^2k^2 + 2Ukm + m^2\right), \nonumber \\
    &= \frac{MK}{12} \left( U^2 (K^2-1) + M^2-1 \right) \epsilon^2 \sin^2\theta \, .
    \label{sum_pm2: appendix}
\end{align}

\noindent Similarly, using Taylor expansion, one can approximate the off-diagonal term as

\begin{align}
    q_{m,k}p_{m,k} &=  \frac{g \epsilon \sin\theta\big(1 - g \epsilon\cos\theta\big)}{1 - 2(g\epsilon)\cos\theta+(g\epsilon)^2}   
     \, , \nonumber \\
    & \approx g\epsilon + (g\epsilon)^2\sin\theta\cos\theta \nonumber \, ,
\end{align}

\noindent Note that, due to symmetry, $\sum_{m,k} g = \sum_{m,k} Uk+m = 0$, hence the summation becomes
\begin{align}
    \sum_{m,k} q_{m,k}p_{m,k} \approx \frac{MK}{12} \left(U^2(K^2-1) + M^2-1 \right) \epsilon^2 \sin\theta\cos\theta \, ,
    \label{q_mp_m: appendix}
\end{align}

\noindent Therefore, the proof of Theorem \ref{Theorem} is completed.

 }


\bibliographystyle{IEEEtran}
\bibliography{Ref}
\end{document}